# Single Classifier-based Passive System for Source Printer Classification using Local Texture Features

Sharad Joshi, Nitin Khanna *Member, IEEE*

*Abstract*—An important aspect of examining printed documents for potential forgeries and copyright infringement is the identification of source printer as it can be helpful for ascertaining the leak and detecting forged documents. This paper proposes a system for classification of source printer from scanned images of printed documents using all the printed letters simultaneously. This system uses local texture patterns based features and a single classifier for classifying all the printed letters. Letters are extracted from scanned images using connected component analysis followed by morphological filtering without the need of using an OCR. Each letter is sub-divided into a flat region and an edge region, and local tetra patterns are estimated separately for these two regions. A strategically constructed pooling technique is used to extract the final feature vectors. The proposed method has been tested on both a publicly available dataset of 10 printers and a new dataset of 18 printers scanned at a resolution of 600 dpi as well as 300 dpi printed in four different fonts. The results indicate shape independence property in the proposed method as using a single classifier it outperforms existing handcrafted feature-based methods and needs much smaller number of training pages by using all the printed letters.

## I. Introduction

WITH the advancement of printing technology, printers have become more affordable, accessible and deployable for the common public. Simultaneously, even in this digital era, printed material remains a widely used form of display for reading general textual information, particularly in developing countries. Further, the usage of paper for storage, display and communication of critical information persists in significant amounts around the globe. For example, according to a survey report released by international data corporation (IDC) in 2012 [?], two million pages are printed in EMEA (Europe, Middle East, and Africa) region every minute. The estimated number of printed pages all over the world is around six million pages per minute. Although the proportional use of digital content such as e-books, e-newspapers and soft copies of articles is increasing rapidly, for some applications, such as legal tenders, printed material is still more suited than their digital counterparts. Especially in under-developed or developing countries, cutting edge and low-cost digital display and reproduction technologies are not so widely available, and thus hard-copy (physical printed page) remains the most common communication medium. Amongst different mechanisms of generating printed media, such as newspaper/magazine printing in a printing press and photocopying; printing files using a desktop printer is one of the most popular ways of creating a printed content due to its low cost and ease of availability.

Use of paper is not only widespread in financial and judicial processes such as lease agreements, purchase invoices/bills, and court judgments but also in personal identity verification at important places such as airports or secure meeting places which may be prone to unfriendly intruders. Such extensive use of printed documents pose serious challenges for law enforcement agencies and makes verification of printed documents a crucial task for providing proper security. Thus, for monitoring documents in bulk, fast and accurate methods for printed document investigation are indispensable. Traditional investigation methods, mainly based on chemical analysis of toner, tend to be slow and at times *intrusive* concerning the investigated document i.e. they may damage the investigated document either partially or entirely.

An important aspect in forgery detection using digital investigation is to ascertain the source of a printed document. Clues about type, brand or model of the printing machine could help in distinguishing the forged documents from huge volumes of printed documents like in the case of counterfeit currencies [1]. For source identification and authentication of printed documents, several methods based on digital image processing have been proposed by researchers. These state of the art techniques work by acquiring printer's device signature [2].

Artifacts, which lead to these signatures are the result of defects in printing technology. Printer manufacturers work on improvements in printing technology by removing any visual artifacts present in the printouts but to make the printers more cost effective, resources are not engaged for minimizing the artifacts which are usually not noticeable by the naked eye. All the experiments presented in this paper have been conducted on text documents printed from laser and ink-jet printers. Working of a laser printer can be summarized into 6 steps: charging, exposition, development, transferring, fusion and cleaning [3]. The signature obtained from the digitally scanned version of any printed document comprises a combination of the artifact(s) that may be caused during one or more of these steps.

Source printer identification problem for a closed-set can This material is based upon work supported by the Board of Research in Nuclear Sciences (BRNS), Department of Atomic Energy (DAE), Government of India under the project DAE-BRNS-ATC-34/14/45/2014-BRNS and Visvesvaraya PhD Scheme, Ministry of the Electronics & Information Technology (MeitY), Government of India.

S. Joshi and N. Khanna are with the Multimedia Analysis and Security Lab, Electrical Engineering, Indian Institute of Technology Gandhinagar, Gujarat, 382355 India e-mail: (nitinkhanna@iitgn.ac.in).

be modeled as a classification problem and thus can be dealt using a suitable classification algorithm such as support vector machines (SVM). Irrespective of the detailed working of methods, most of the methods proposed so far in the literature have experimented with a single language script environment i.e. only those documents can be investigated which contain the same letter or group of letters that were used for generating the training model [3]–[6]. The major exceptions to this are the geometric distortion based methods [7]–[9] which are content independent. However, currently, their performance suffers due to limitations on accurate ground truth generation. Also, a recent work [10] illustrates a text independent technique via synthetic texture generation. Though it does not explicitly mention usage of multiple languages or fonts in any of the experiments but it mentions the use of random text for their experiments. Apart from the disparity in language, the features proposed so far also depend heavily on the font type and font size of the alphabets. The proposed method aims to address these limitations by using features from the local texture patterns present in a printed character or symbol. For the purpose of this research, all printed letters are simultaneously used for feature extraction and classification. This method stands on the hypothesis that each printer induces unique artifacts in the printed document irrespective of the content of that document. So, even if the font or size of the letters changes, the way a printer treats a horizontal texture or a curved texture of certain shape remains the same. With this objective, various operators for estimating local texture patterns [11]–[14] are tested. Out of those, the local tetra patterns (LTrP) [14] gives the highest classification accuracy. It is worth noting that the scope of this hypothesis is limited to text documents which contain only two levels of toner ink i.e. black and white. On the other hand, a printed image generally tries to approximate continuous-tone ink using half-tones, which can have multiple shades of gray in addition to black and white. Major highlights of the system proposed in this paper are:

- A first of its kind, single classifier based approach which surpasses limitations of steps involved in identifying each letter of a language such as using optical character recognition (OCR).
- Requires lesser number of pages for training in comparison to existing methods because all the letters are used while training a single classifier.
- A strategic feature extraction process that involves dividing each letter into three regions along with Post-Extraction Pooling (PoEP).
- A method which outperforms all the existing methods based on handcrafted features and closely matches the performance of compute-intensive data-driven state-of-art.
- Effective at 300dpi resolution as well while the existing methods for printer classification have been tested only at 600dpi or higher.
- Supports faster scanning and lesser processing requirement thus has potential for processing large volumes of printed documents in public and private offices, courts and other places of mass usage of paper.

## II. Methods of Source Printer Identification

Source printer identification from a printed document via digital techniques has been researched extensively in the last decade [3]–[7], [9], [15]–[20]. Digital methods for source sensor classification can be broadly classified into two categories: Extrinsic and Intrinsic [21], [22].

The first category of digital methods comprises of external signature based methods in which some external signature is embedded by manipulating some characteristics and/or parameters of the document that may be related to the printer mechanism. In this paper, external signature added before or after the printing process is termed as watermark [23] while external signature added during the printing process is termed as extrinsic signature [22]. In this sense, extrinsic signatures are more robust than watermarks because, to tamper the former, printer mechanism needs to be modified. A prominent technique to embed an extrinsic signature is via modulation of half-tones. Half-toning is an integral part of printing gray level images by electrophotographic printers. Half-tone signatures have been analyzed in the Fourier domain. In particular, use of laser intensity modulation to modulate the dot size of halftones in printed images is illustrated in [24]. However, half-toning is used only for gray level images and not for black and white text documents. Though the extrinsic methods are promising, they require the use of sophisticated equipment and technology to manipulate the printing process.

### A. Source Classification using Intrinsic Signatures

The second category consists of the intrinsic signature based methods which rely on measuring printer specific artifacts induced during the printing process. These intrinsic signatures may result from minor imperfections in the mechanical parts such as optical photoconductor (OPC) drum and gear mechanisms [25] or other internal processes involved in printing that are specific to certain make and model of printers.

*1) Early Methods:* Half-toning based features were one of the earliest to be utilized as an intrinsic signature for identifying the source of a printed document [18], [26], [27]. However, it is a characteristic of printed images that require mimicking different gray scales. On the other hand, printed text documents require only two gray scale levels, so no halftoning is used. In addition to half-toning, another phenomenon that was used for identifying a printer was banding [15] which refers to non-uniform bright and dark patches that appear in a printed document perpendicular to the process direction. When tested on 40 letters each from 5 different printers, this method gives 100% classification accuracy for 3 of them. These approaches, though promising, require a scanned input of very high resolution (2400 dpi and higher). While, at lower resolutions (300-1200 dpi), the banding effect can be perceived as the non-uniform texture in printed letters combined with the spread of ink dots. Consequently, this led to the development of texture based analysis.

*2) Texture Based Methods:* In [3] 22 features based on gray-level co-occurrence matrices (GLCM) were used for printer classification using random text documents printed from 10 laser printers and scanned at 2400 dpi and 8 bits/pixel in grayscale. From these scanned documents, they extract a single type of letter ('e'). From a total of 10 printers with 300 e's each, barring one, rest all printers are classified correctly based on majority voting. On similar lines, a method has been proposed for documents printed in the Chinese language [4]. It extracts 22 GLCM features and 12 discrete wavelet transform (DWT) based features from a single Chinese letter. This method reports maximum average classification accuracy of 98.64% with 12 printers examined with two types of fonts and four different font sizes. The accuracy was better than the accuracy obtained from only GLCM features on the same dataset.

Use of multi-dimensional and multi-scale GLCM features for printer classification was proposed in [5]. This method applied the scheme on selected individual letters (only e's), a portion of a document (termed as 'frames') as well as the whole document. Unlike previous techniques, while using 'frames' and the whole document this method used all the letters present in the frame/document. They used SVM as a classifier and the highest average page-level accuracy reported by them is 97.60% based on majority voting on the individual classification results of all letters in a single page. But, these results are only with e's of same size which has been discussed by the same authors in [6]. Same authors have presented the advantages of using letters over frames and concluded that letters are better suited in many practical scenarios [6]. Their recent work [6] is based on convolutional neural networks (CNNs) which trains a group of three separate networks in parallel with inputs as e's, the median residual of e's and average residual of e's. Another group of three networks is trained with the above settings using letter 'a'. So, this is also a multi-classifier approach which gives 97.33% page-level accuracy on the same dataset [5]. However, they report that the letter extraction used by them in both these papers is not very accurate for letters other than 'e' and 'a'. Also, these two approaches require generation of reference letter of same size and font for letter extraction process. Moreover, the second approach use a fixed size of letter as input, $28 \times 28$ to be precise. So, they are prone to changes in letter size which may occur in practical scenarios due to multiple size and scaling options available during printing.

Some of the proposed systems also use texture patterns at micrometric scale [20]. In addition to printer classification, a number of texture-based systems are proposed for classifying various printing techniques [28]–[30] and for forgery detection [31]–[33]

*3) Geometric Distortion Based Methods:* Apart from the texture based methods, there is a comparatively newer category of methods that relies on the fact that in each document there is some printer induced geometric distortion in the form of translation, rotation, and scaling distortions [7]–[9]. This category of methods are particularly robust to changes in toner levels and are more or less independent of text content. But as of now, there is no technique that provides a framework for letter-level printer classification without the use of a reference image. This limitation can be addressed successfully by texture based methods.

A critical limitation of the texture based methods discussed so far is that their performance deteriorates significantly with text containing different fonts, sizes, and languages. In this paper, we try to investigate the performance of existing methods with four different types of fonts. Though, an exception to the above limitation is a recently proposed method [10] which computes synthetic printer textures from stable fragments within a letter based on neighboring pixel values. This method reports the highest page level accuracy of 96.67% with experimentation on a dataset containing random text obtained from 12 printers. In sharp contrast, the proposed method extracts features based on the real texture information and does not require computation of synthetic textures. Also, the accuracies of classification, reported in this paper, are for a group of letters instead of the whole page and thus the proposed method can be generalized for forgery detection as well.

### III. LOCAL PATTERNS

Local binary patterns (LBP) were first introduced by Ojala et.al [11] as a measure of texture present in grayscale images and used for texture classification. These original patterns are invariant to linear transformations in intensity level or gray scale but not invariant to rotation. This initial formulation of LBP was further extended as rotation invariant local binary patterns (RILBP) to make them invariant to rotation. Further, the authors identified a set of more frequently occurring patterns and termed them as uniform local binary patterns (ULBP) based on the condition that they had at most two transitions between the binary digits 0 and 1, when the patterns were viewed as circular 8-bit strings. These uniform patterns denote edges, corners, and spots in an image. When eight neighbors are used to estimate local binary patterns, for original LBP there are 256 possible patterns while for ULBP although this number reduces to 58, still these 58 patterns together represent nearly 90 percent of all the patterns present in an image [11].

There may be gray scale variations induced during printing due to variations in toner levels and also during scanning due to variations in light exposure. So, for the texture based printer classification approach to be successful, used texture measure or feature should be invariant to linear transformations of intensity levels. This makes LBP operator a good candidate for feature extraction step used in a printer classification system as with gray scale invariance they can provide significant robustness against toner level variations. The gray scale invariance of LBP is attributed to the fact that it uses signs of differences between the center pixel intensity and intensity of pixels in the neighborhood to estimate texture pattern for each pixel. Thus, if the intensity value of a center pixel and its neighborhood is similarly scaled, the sign of difference between the center pixel intensity and intensity of pixels in the neighborhood will remain same and so will be the LBP derived features. LBP operator has been successfully used in the literature for a variety of other applications as well such as face recognition [34] and palm-print identification [35].

LBP operator for a $3 \times 3$ neighborhood is defined as follows:

$$LBP(p) = \sum_{n=0}^{7} 2^n u[I(q_n) - I(p)] \quad (1)$$

Here, $p$ and $q_n$ denote locations of central pixel and $n^{th}$ neighboring pixel, respectively. The ordering of neighboring pixels does not affect final classification as long as one maintains consistency across all the samples. So, the starting index, $n = 0$, can be selected as any one of the eight neighbors and then rest of the neighbors can be sequentially traversed in clockwise or anticlockwise direction. Weights for the binary levels are given by $u[I(q_n) - I(p)]$ where, $u[.]$ denotes the unit step function. These patterns are further used to estimate discrete occurrence histograms which act as powerful features because they combine statistical and structural properties of an image.

LBP utilizes a single value threshold to differentiate between the two binary levels and thus it can be sensitive to noise, particularly in flat areas of an image. To tackle this problem, the LBP patterns were modified to a generalized form resulting in the local ternary pattern (LTP) for solving face recognition problem [12]. LTP utilizes a broader boundary to differentiate between the two levels and simultaneously it also utilizes this boundary to delineate a third level for computing the pattern. So, for LTP the weights of LBP (role played by unit step function, $u[.]$, in Equation 1) get modified to the following:

$$w_{LTP}[x] = \begin{cases} 1, & x \geq T \\ 0, & |x| < T \\ -1, & x \leq -T \end{cases} \quad (2)$$

where the threshold $T$ controls transition width for differentiating between two levels. This is a three-level pattern and hence the name local ternary pattern (LTP). Next, the authors proposed to split this pattern into upper and lower patterns. At last, similar to the LBP case, occurrence histograms are computed separately for lower and upper patterns. Another prominent LBP variant is the local derivative pattern (LDP) which has been used for face recognition [13]. It is a non-directional higher order extension of LBP, treating LBP as the first order pattern.

In the above variants, the emphasis is solely on the difference in intensities, but to capture more information, gradient direction of the center pixel may also provide important clues. This has been captured by local tetra patterns (LTrP) [14]. This four direction code was earlier used for content-based image retrieval. It's first step involves mapping the direction corresponding to the central pixel onto one of the four possible directions using following equation:

$$G_{dir}(p) = \begin{cases} 1, & G_h(p) \geq 0 \text{ and } G_v(p) \geq 0 \\ 2, & G_h(p) < 0 \text{ and } G_v(p) \geq 0 \\ 3, & G_h(p) < 0 \text{ and } G_v(p) < 0 \\ 4, & G_h(p) \geq 0 \text{ and } G_v(p) < 0 \end{cases} \quad (3)$$

Here, for horizontal and vertical neighbors $q_h$ and $q_v$ of central pixel $p$, gradients are given by $G_h(p) = I(q_h) - I(p)$ and $G_v(p) = I(q_v) - I(p)$, respectively. Based on this, for each $3 \times 3$ window, the 8-bit LTrP code at the central pixel $p$ is estimated as:

$$\begin{aligned} LTrP(p) = \{&w_{LTrP}(G_{dir}(p), G_{dir}(q_0)), \\ &w_{LTrP}(G_{dir}(p), G_{dir}(q_1)), ..., \\ &w_{LTrP}(G_{dir}(p), G_{dir}(q_7))\}, \end{aligned} \quad (4)$$

where,

$$w_{LTrP}(G_{dir}(p), G_{dir}(q_n)) = \begin{cases} 0, & G_{dir}(p) = G_{dir}(q_n) \\ G_{dir}(q_n), & else \end{cases} \quad (5)$$

Next, this tetra pattern is converted into 3 binary patterns corresponding to 3 non zero directions by replacing the corresponding direction's mapped value by ones and replacing all other mapped direction values to zeros, thus, producing 12 patterns ($4 \times 3$, 4 possible values of $G_{dir}(p)$ and for each of them 3 binary patterns) [14]. In addition, a magnitude based pattern is also introduced which is given by:

$$M(p) = \sum_{n=0}^{7} 2^n \times u[G_m(p) - G_m(q_n)] \quad (6)$$

where, the gradient magnitude $G_m(.)$ is estimated as $G_m(x) = \sqrt{(G_h(x))^2 + (G_v(x))^2}$. A more detailed discussion of LBP and its different variants is provided in [36]. The system proposed in this paper uses LTP-based features for printer classification. The superior performance of LTP-based features is evident from experiments performed on other variants of LBP.

## IV. PROPOSED METHOD

The main goal of this paper is to propose features of printed text documents which capture essential characteristics of printer's texture related intrinsic signatures and are shape independent. Figure 1 depicts an overview of the proposed method for source printer identification of printed document. The input to the proposed system is a hardcopy or printed document. Thus, we are free to choose any suitable scanner and scanning parameters (such as scanning resolution) appropriate for further digital image processing, as long as we use the same setup for scanning all the documents used to train the system and test documents of unknown origin. The first step in the proposed system is to obtain a digital image of the document to be investigated, by scanning it using a reference scanner $S$. This scanned image (digital version of the hardcopy or printed document) is used as input for further steps.

### A. Pre-processing

The scanned image is passed through a pre-processing stage before extracting any features for source printer classification. This step consists of an optional step of cropping some portion of the image margins. This step removes any effect of printing noise which dominates at the borders of the page. This cropping is optional because most printers introduce some inherent empty margin space. Since, the proposed method does not require OCR which is sensitive to small rotations of a document. Therefore, in contrast to other similar methods [3], [4], the proposed method works without any rotation correction and is robust to small rotations of a document.

### B. Feature Extraction

Using the output from preprocessing stage, the final feature vectors are extracted in the following manner:

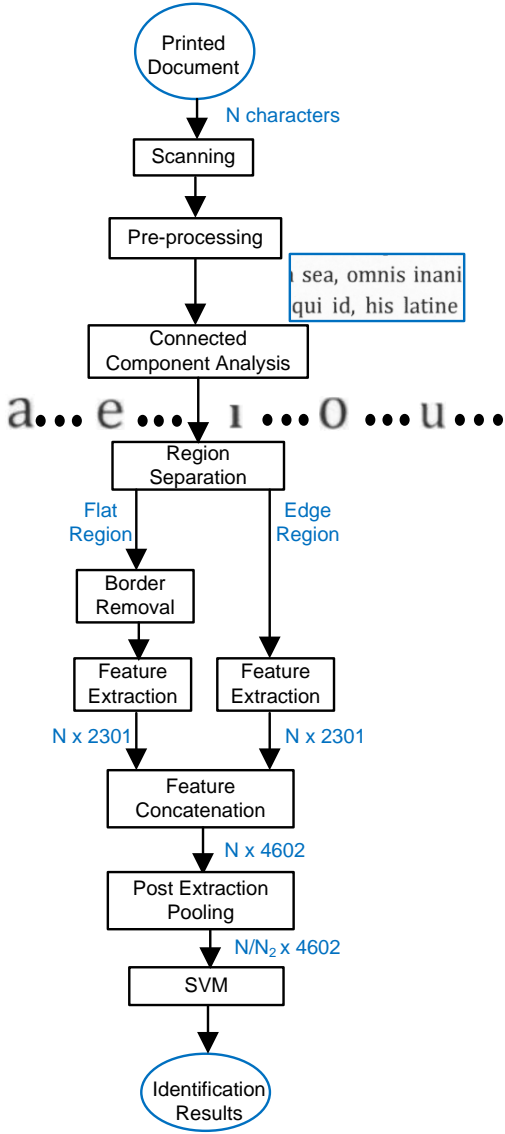

Fig. 1: Block Diagram of the Proposed Approach.

estimated and averaged using a mean filter of size five so that it can be well approximated as a bimodal histogram. The average intensity ($\mu$) is computed from intensities corresponding to the two peaks in the histogram. Two thresholds are empirically chosen as $\alpha\mu$ and $\beta\mu$ so as to divide each letter into three regions: a flat region ($F$), an edge region ($E$) and a background region ($B$). For the experiments reported in this paper, $\alpha$ and $\beta$ are empirically selected as 0.71 and 1.52 respectively. The background region, which should not ideally contain any ink, will have other types of noises such as scanning noise. This can be inferred by noticing the zoomed versions of letter 'A' (Figure 2). So, the proposed method leaves out the background region as it is highly contaminated by noises which are not printer specific. Thus, features are extracted separately from the remaining two regions.

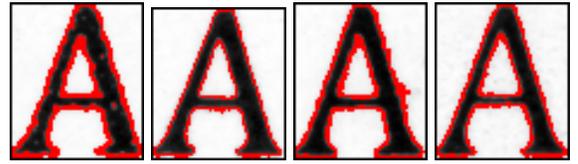

Fig. 2: Zoomed version of a letter 'A' printed from four different printers from our dataset: LB1, LC1, LC4, and LC9 (Table I). Black region corresponds to flat area F and red region depicts the edge area E. The remaining portion is background region.

*2) Feature Vector Formation:* LBP variants have been shown to work better in combination with Gabor filtering [36]. So, first of all, each letter image (both edge and flat regions) is subjected to Gabor filtering using a Gabor filter bank of three scales and two orientations ($0°$ and $90°$) in the same way as in [14]. Gabor filter size is fixed at $10 \times 10$. Then, border pixels are removed from the flat region. Here, border pixels are defined as any pixel in the flat region which has at least one neighbor in the edge region considered over a neighborhood of size $3 \times 3$. Removal of border pixels makes the feature extraction process inherently shape-independent. Since edge region contains a very small number of pixels as compared to the flat region with an average width of one pixel, so border removal process is not performed for the edge region. Next, Equations 3, 4, 5 and 6 of the LTrP operator are evaluated separately on each pixel of edge region $E$ and flat region $F$.

Now, one possibility is to train $c$ classifiers, each trained separately using feature vectors obtained from training instances of a particular letter. Here $c$ represents the number of letters in the character set of a particular language. For example, while working on documents printed in English, $26 \times 2$ classifiers will be required. However, given the availability of several different font types and font sizes, it would further increase the value of $c$ and in turn, training complexity. Also, printed documents have been known to be affected by aging, so the document under test needs to be compared against a training data obtained around the same time. Thus, in several scenarios, it might be difficult to find sufficient training data satisfying the aforementioned conditions. Another issue with this procedure is that even if we can afford to use separate classifiers in

*1) Region Separation:* In this step, a scanned image is subjected to connected component analysis. Resulting components are further filtered to remove components of spurious size and orientation. This is done by setting upper and lower thresholds on component area. With experiments on our dataset, upper threshold is set as 4 times the median of areas covered by all the components on a single page. On the other hand, lower threshold is set to 0.5 times the median of areas. This gives good accuracy in separating individual letters in different fonts of English language. The nature of texture inside a letter's boundary and at the boundary edge are very different. Thus, it is more appropriate to divide the entire region inside the bounding box corresponding to a particular letter into multiple parts. This separation into multiple parts is done with the help of intensity histogram corresponding to different bounding boxes. A similar division of regions is utilized in [30] for classifying printing techniques (inkjet vs. laser). Histogram of the image portion corresponding to each bounding box is some cases, such a solution would require the use of OCR to ascertain the identity of a letter being tested. However, OCR's accuracy is not uniformly good for all languages and fonts. It is also dependent upon the rotation induced during printing and scanning. For example, Matlab's inbuilt OCR highly misclassifies English letters in tilted documents. Even if a tilt correction procedure is applied to correct the induced rotation, the accuracy of OCR will depend upon the efficiency of tilt correction which would have a direct impact on the final classification accuracy. To tackle these challenges, we propose the use of a common classifier. Moreover, since the proposed method does not utilize a letter's shape information, this common classifier is expected to work well. This means that the language, font types and font sizes used in a text document does not impact the number of classifiers required.

From the output of the LTrP operator, uniform patterns are selected and a $4602$ ($13 \times 59 \times 3 \times 2$ - 12 patterns corresponding to Equation 5 and one corresponding to Equation 6) bin histogram of the selected patterns is estimated from a letter corresponding to each region for each of the three scales corresponding to Gabor filter. The number of bins corresponds to the maximum number of possible uniform patterns plus one bin for all the non-uniform patterns. As per our assumption, the value of each bin in this histogram represents the probability of occurrence of that pattern in that region ($F$ or $E$) of a particular letter.

*3) Post-Extraction Pooling (PoEP):* Before the classifier is trained, feature vectors from $N_2$ letters are pooled into a single average or pooled feature vector. This step compensates for any unwanted noise or distortions as this pooled vector is a representative value from multiple letters. If we represent each feature vector corresponding to a training sample by a single point in a high dimensional space, then all the feature vectors of the same class would form a cloud in that space. It is assumed that before the pooling operation, all such points are randomly distributed so that the overall effect would be to remove outliers (noise-ridden samples) or to tighten the cluster corresponding to the cloud of points from a particular printer. PoEP is superior to majority voting procedure since the latter considers only the labels of individual sample points (corresponding to their feature vector) and chooses the most occurring label, whereas, the former approach computes mean of sample points based on their relative distances from each other. So, this setting considers the orientation of sample points about the center. The distribution of these orientations is directly dependent on the distribution of unwanted noise. As a result, if we assume the unwanted noise to be randomly distributed across letters on a single page then, PoEP can better utilize the randomness of this noise to compensate for it. Moreover, the assumption that the data points are randomly distributed is satisfied by applying PoEP on consecutively printed letters which were randomly picked during dataset preparation and by choosing a large enough value of $N_2$. PoEP is applied only across letters extracted from the same page as classifier's final decisions should correspond to letters occurring on a single page and not across multiple pages. So, each page should be independently processed. Depending on the type of application, PoEP can be generalized for any method which extracts feature vectors for any learning procedure.

The problem of close-set printer identification can be modeled as a classification problem. This paper uses one of the most widely used supervised classification technique in multimedia forensics, support vector machine (SVM) (LIBSVM library [37]) for classifying feature vectors obtained from each group of $N_2$ letters. Final decisions about source of a printed page are obtained by taking majority voting over all the groups of $N_2$ letters present on that page.

## V. Experimental Results and Discussions

This section outlines the dataset used, types of experiments performed along with results and comparison with existing methods.

TABLE I: Details of printers used in experiments.

| Printer ID | Printer Brand | Printer Model | Printer Resolution (in dpi) |
|---|---|---|---|
| LB1 | Brothers | DCP 7065DN | $2400 \times 600$ |
| LC1 | Canon | D520 | $1200 \times 600$ |
| LC2 | Canon | I6570 | $2400 \times 600$ |
| LC3 | Canon | IR 5000 | $2400 \times 600$ |
| LC4 | Canon | IR 7095 | $1200 \times 600$ |
| LC5 | Canon | IR 8500 | $2400 \times 600$ |
| LC6 | Canon | LBP 2900B | $2400 \times 600$ |
| LC7 | Canon | LBP 5050 | $9600 \times 600$ |
| LC8 | Canon | MF 4320 | $600 \times 600$ |
| LC9 | Canon | MF 4820d | $600 \times 600$ |
| LC10 | Canon | MF 4820d | $600 \times 600$ |
| LC11 | Canon | MF 4820d | $600 \times 600$ |
| IE1 | Epson | L800 | $5760 \times 1440$ |
| IE2 | Epson | EL 360 | $1200 \times 600$ |
| LH1 | HP | 1020 | $600 \times 600$ |
| LH2 | HP | M1005 | $600 \times 600$ |
| LK1 | Konica Minolta | Bizhub 215 | $600 \times 600$ |
| LR1 | Ricoh | MP 5002 | $600 \times 600$ |

### A. Dataset and General Experimental Settings

Since the only publicly available dataset in the English language has documents printed in a single font from only ten laser printers [5], and another publically available dataset is in the German language with 400 dpi pre-processed (skew correction and border noise removal) scanned documents [38]. Therefore, in addition to evaluating the performance of the proposed system on existing dataset [5], we have also prepared a new dataset consisting of 720 pages printed from eighteen printers (Table I) to test the efficacy of the proposed system on a larger database containing multiple fonts. All these pages contain random letters of English language (generated from Lorem ipsum [39]) in four different fonts. From each printer, there are twenty-five pages in Cambria font while five pages each are in Arial, Comic Sans, and Times New Roman fonts. Pages in Arial font have font size 11, while pages in other three fonts have font size 12 as per the general settings used in most legal documents. Cambria, Arial and Times New Roman are included here as they are widely used in most legal documents while Comic Sans is included because it looks quite different from the rest of the fonts. In our dataset, the number of pages per printer is smaller than existing dataset as our method uses all letters printed on a page and also we have reported accuracies on group of letters. All these printed pages are scanned using a single reference scanner (Epson Perfection V600 Photo Scanner) at 600 dpi and 300 dpi resolutions. There are three printers which are of the same brand and model (LC9, LC10, and LC11 in Table I). Due to the closeness of their printer signatures, the intra-model scenario requires separate analysis. So, the initial experiments on our dataset for optimizing system parameters have been performed only on documents obtained from printers of different models (16 printers in Table I except LC10 & LC11), while the final experiment includes all the 18 printers. The performance of the proposed method is comprehensively analyzed by conducting experiments with a larger number of printers as compared to the existing public dataset. Experiments are performed on four different fonts and at two different scanning resolutions. The general pipeline for these experiments stays the same as shown in Figure 1, while the feature extraction block is modified in different experiments (Sections V-B to V-E).

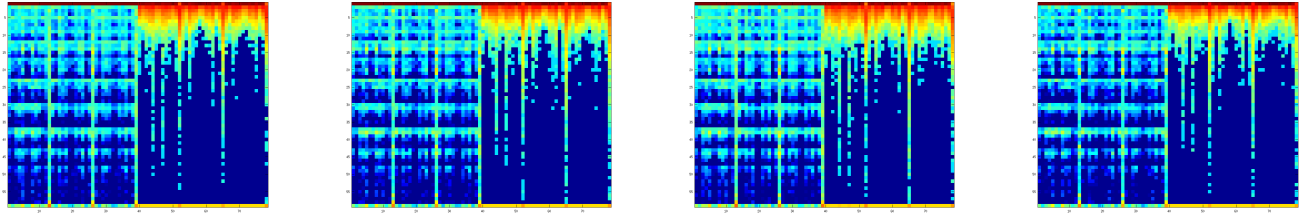

Fig. 3: Colormap depicting feature vectors reshaped to size $59 \times 78$ over $N_2 = 40$ letters printed from LB1, LC1, LC4, and LC9 printers of our dataset (Table I). Here each column corresponds to a single histogram.

The proposed method is compared with state of the art methods (Section II) based on GLCM [3], DWT [16], GLCM-DWT [4], Cross Center-symmetric LTP (CCS-LTP) [40], multi-directional GLCM (GLCM_MD) [5] and multi-directional multi-scale GLCM (GLCM_MD_MS) [5]. Here, convolution texture gradient filter (CTFG) [5] has not been included as a baseline method for comparison as the proposed method has been tested on letters and GLCM_MD and GLCM_MD_MS have been shown to outperform CTGF on letters [5].

### B. Performance of LBP Variants and Existing Methods

These experiments aim to identify the most suitable LBP variant for printer classification task by using different LBP variants for feature extraction while keeping all other parameters of the system fixed. For comparing the efficacy of the proposed method with the existing methods, the settings of these experiments are kept same as in most existing methods, such as using only **e's** and extracting a single feature vector from each letter without any region separation and PoEP [3], [5]. From each of the **e's** occurring in a page, features are extracted using ULBP, RILBP, LTP, local derivative pattern (LDP) [13], local tetra patterns (LTrP) [14] (Section III), Gabor rotation invariant LBP (GRILBP), Gabor LTP (GLTP), Gabor complete LBP (GCLBP), and Gabor uniform LBP (GULBP) operators (Section IV-B2). Existing texture-based methods for printer classification (Section II) are also subjected to exactly same experiments as far as training, and testing sets are concerned. Twenty-five pages (approximately 300 e's on each page) printed from each printer in Cambria font are used in this set of experiments. Out of these, letters extracted from randomly selected twenty pages are used for training the classifier and the rest for testing. Table II summarizes the results of experiments comparing different LBP variants by reporting average accuracy across all 16 printers for classifying an individual letter **e**. Since the GLTrP based method outperforms the rest of the LBP variants with 96% average classification accuracy (Table II). Therefore, for rest of the experiments related to the proposed system, we will be using GLTrP for feature extraction. The proposed system using GLTrP for feature extraction and under similar experimental settings as reported by existing methods performs slightly better than the existing baseline methods as it has 96% average classification accuracy while GLCM_MD_MS based method gives 95.5% average classification accuracy, highest amongst all the existing methods (Table III).

TABLE II: Performance of LBP variants using only e's from 16 printers of our dataset.

| Feature Extraction Method | Feature Size (1-D) | Classification Accuracy (%) |
|---|---|---|
| ULBP | 1 x 59 | 79.1 |
| RILBP | 1 x 37 | 61.5 |
| LTP | 2 x 59 | 75.9 |
| LDP | 4 x 59 | 68.4 |
| LTrP | 13 x 59 | 83.1 |
| GULBP | 1 x 59 | 90.0 |
| GRILBP | 1 x 37 | 68.9 |
| GLTP | 2 x 59 | 82.7 |
| GLDP | 4 x 59 | 91.7 |
| GLTrP | 13 x 59 | **96.0** |

TABLE III: Results of existing methods using only e's from 16 printers of our dataset.

| Feature Extraction Method | Feature Size (1-D) | Classification Accuracy (%) |
|---|---|---|
| DWT | 1 x 12 | 86.2 |
| GLCM | 1 x 22 | 90.6 |
| DWT+GLCM | 1 x 34 | 93.4 |
| GLCM_MD | 1 x 176 | 93.0 |
| GLCM_MD_MS | 1 x 704 | **95.5** |
| CCS LTP | 1 x 128 | 57.8 |

TABLE IV: Average classification accuracy using all letters and different values of $N_2$.

| # Train Samples | 3665 | 1828 | 1215 | 910 | 726 |
|---|---|---|---|---|---|
| # Test Samples | 38580 | 19328 | 12914 | 9703 | 7789 |
| $N_2$ | 10 | 20 | 30 | 40 | 50 |
| DWT | 88.1% | 88.4% | 90.5% | 85.0% | 90.7% |
| GLCM | 86.1% | 86.1% | 86.7% | 87.3% | 87.0% |
| DWT+GLCM | 91.8% | 92.4% | 92.6% | 92.3% | 89.2% |
| GLCM_MD | 86.6% | 87.4% | 87.6% | 86.6% | 87.0% |
| GLCM_MD_MS | 88.2% | 87.8% | 89.3% | 88.5% | 89.7% |
| Proposed | 98.6% | 99.2% | 99.3% | 99.4% | 98.6% |

### C. Effectiveness of Post Extraction Pooling (PoEP)

These experiments are conducted to investigate the effect of PoEP in combination with region separation with same font data (Cambria) in training and testing. In this set of experiments, all small and capital letters are extracted using connected component analysis, and each of them is subdivided into flat $F$ and edge $E$ regions followed by PoEP with different values of $N_2$ (Section IV). GLTrP and existing methods (Section II) are applied separately on each sub-region of all the letters. Table IV shows the average classification accuracies of different methods for different values of $N_2$ when all the letters are used for classification. First two rows in Table IV show the total number of samples/feature vectors (corresponding to 16 printers) fed in the classifier for training and testing, corresponding to mutually exclusive randomly chosen sets of 1 and 10 pages, respectively. Note here that as $N_2$ goes from 10 to 50, the actual number of training samples decreases but the effective total number of letters used to train the classifier remains the same and same is true for test samples. Thus there is a trade-off between smaller and larger values of $N_2$, with an increase in the value of $N_2$, individual feature vectors fed to the classifier become more shape independent, but the classifier will have lesser number of feature vectors to build the model. The average classification accuracy for the proposed method increase when $N_2$ is increased from 10 and out of the five values of $N_2$, $N_2 = 40$ gives the highest average classification accuracy of 99.4%. Hence for rest of the experiments reported in this paper, PoEP is used with $N_2 = 40$. Further, the proposed method consistently outperforms the existing methods by a significant margin for all the values of $N_2$, ranging from 10 to 50 (TableIV). For example, if a feature vector is formed from a group of 10 letters, the proposed method gives an average classification accuracy of 98.6% while DWT+GLCM based method gives an average classification accuracy of 91.8%, the highest amongst all the existing methods. Results summarized in TableIV show that the proposed method satisfies shape independence property to a greater extent and will allow us to build a single classifier for all the letters printed on a page. It will not only decrease the number of pages required for training a classifier but also alleviate the problems associated with using OCR to classify the letters.

The effect of PoEP technique is further examined by visualizing the distribution of features in a lower dimensional space. This is done by extracting features from training samples for different $N_2$ and applying linear discriminant analysis (LDA) to obtain reduced dimensional feature vectors. The projected dimensions corresponding to the top two eigenvalues are plotted in Figure 4. These plots indicate that PoEP technique in the proposed method reduces intra-class separation and increases inter-class separation. Thus, different clusters corresponding to different printers (shown in different colors in Figure 4) have much lesser overlap in Figure 4(b) as compared to Figure 4(a).

TABLE V: Page level confusion matrix for *CNN-$\{S^{raw}, S^{med}, S^{avg}\}_{a,e}$* [6], *CTGF-GLCM-MD-MS* [6] and the proposed *CC-RS-LTrP-PoEP* on the existing dataset [5] corresponding to $5 \times 2$ cross-validation folds of [6].

| | B4070 | C1150 | C3240 | C4370 | H1518 | H225A | H225B | LE260 | OC330 | SC315 |
|---|---|---|---|---|---|---|---|---|---|---|
| CC-RS-LTrP-PoEP | 99.61 | 97.26 | 98.54 | 98.67 | 94.99 | 93.47 | 91.95 | 98.35 | 98.35 | 100 |
| CTGF-GLCM-MD-MS | 98.67 | 99.28 | 97.83 | 98.50 | 86.83 | 96.98 | 87.10 | 98.66 | 100 | 99.17 |
| CNN-$\{S^{raw}, S^{med}, S^{avg}\}_{a,e}$ | 99.50 | 99.48 | 98.83 | 100 | 89.17 | 93.10 | 93.45 | 99.50 | 100 | 100 |
| Miss-classification (CC-RS-LTrP-PoEP) | 0.39 (C3240) | 1.20 (B4070) | 0.96 (C4370) | 1.00 (SC315) | 3.81 (C1150) | 6.53 (H225B) | 7.47 (H225A) | 0.99 (C1150) | 0.99 (H225B) | |
| Miss-classification (CTGF-GLCM-MD-MS) | 1.00 (C3240) | 1.72 (C4370) | 2.17 (C4370) | 0.50 (C1150) | 10.33 (H225B) | 2.52 (H225A) | 12.90 (H225A) | 0.67 (C1150) | | 0.50 (C1150) |
| Miss-classification (CNN-$\{S^{raw}, S^{med}, S^{avg}\}_{a,e}$) | 0.33 (C3240) | 0.52 (B4070) | 0.67 (C3240) | | 10.50 (C1150) | 6.90 (H225B) | 6.37 (H225A) | 0.33 (H225A) | | |

Further evaluation of the proposed method is done on the dataset provided in [5] which is the only publicly available printer dataset of English pages. A direct comparison with the results reported in the latest paper on printer forensics [6] is possible because the authors in [5], [6] have made their dataset as well as code publicly available. Their code also reports the exact folds on which they have reported the confusion matrices, Tables XI and XII in [6], showing the best performance obtained by data-driven features and hand-crafted features respectively. Therefore, we have used our proposed system also on the same folds, and the comparison of classification accuracies is reported in Table V. The state-of-art system based on data-driven features *CNN-$\{S^{raw}, S^{med}, S^{avg}\}_{a,e}$* [6] uses a combination of e's, and a's along with their median and average residuals and feeds 1500 dimensional features for every a and e on a page, to SVM classifiers (one SVM classifier for e and another SVM classifier for a). It then takes majority voting on all the decisions corresponding to a page to decide the page-level accuracy (Tables XI in [6]). Similarly, the state-of-art system based on hand-crafted features *CTGF-GLCM-MD-MS* reported in [6] feeds 2999 dimensional features for every e on a page to a SVM classifier and then takes majority voting on all the decisions corresponding to a page to decide the page-level accuracy (Tables XII in [6]). To maintain similarity in testing the proposed method, features are extracted from e's and a's and not from all the characters. Thus, instead of using the optimal choice of $N_2 = 40$ reported in Table IV using all the letters, for this experiment we have performed PoEP on all a's and e's extracted from a single page. This results in great reduction in computational complexity of the proposed method *CC-RS-LTrP-PoEP* as compared to state-of-art systems *CNN-$\{S^{raw}, S^{med}, S^{avg}\}_{a,e}$* and *CTGF-GLCM-MD-MS* because a single 1534 dimensional feature vector (corresponding to $13 \times 59$ dimensional LTrP features for both edge and flat regions) is fed to a SVM classifier for every page. As these experiments only use two letters in the proposed method so good accuracies are obtained at lesser time complexity by avoiding the use of Gabor filter (Section IV-B2). Table V shows the comparison of classification accuracies for each class (in %) obtained by these three methods. For quick comparison classification accuracies of state-of-art systems *CNN-$\{S^{raw}, S^{med}, S^{avg}\}_{a,e}$* and *CTGF-GLCM-MD-MS* are reproduced here from Tables XI and XII in [6]. Classification accuracies of the proposed system *CC-RS-LTrP-PoEP* on same folds of training and testing sets are also reported in Table V. Since for all three methods, misclassification is very small and limited to couple of classes, the $10 \times 10$ confusion metrics are very sparse. Therefore, instead of showing three different confusion metrics of size $10 \times 10$, only the correct classification percentage and for every printer the class corresponding the highest misclassification is shown in Table V. For example, for the proposed method 93.47% pages of printer $H225A$ are correctly classified while 6.53% of this printer's pages are misclassified to printer $H225B$. Following are some of the prominent observations from Table V:

- All three methods *CNN-$\{S^{raw}, S^{med}, S^{avg}\}_{a,e}$*, *CTGF-GLCM-MD-MS* and *CC-RS-LTrP-PoEP* give similar average classification accuracies 97.33%, 96.26% and 97.12%, respectively. Computing resource intensive data-driven *CNN-$\{S^{raw}, S^{med}, S^{avg}\}_{a,e}$* performs slightly better then the systems based on hand-crafted features while the proposed system gives higher accuracy than any other hand-crafted features based system reported in literature [6].

- The minimum accuracy for classifying an individual printer is for $H1518$ using *CNN-$\{S^{raw}, S^{med}, S^{avg}\}_{a,e}$* as well as *CTGF-GLCM-MD-MS* and in both cases this printer primarily gets misclassified to printer $C1150$. These two printers are of different make and model. For example, *CNN-$\{S^{raw}, S^{med}, S^{avg}\}_{a,e}$* has a minimum accuracy for classifying an individual printer as 89.17% for $H1518$, 10.50% pages of printer $H1518$ are predicted to be from printer $C1150$. While the minimum accuracy for classifying an individual printer is for $H225B$ using *CC-RS-LTrP-PoEP* . This minimum value is 91.95%, and a major part of misclassified pages are mapped to another printer of same make and model, $H225A$. For classifying the printer $H1518$, the proposed method gives classification accuracy of 94.99%, and only 3.81% of its pages are misclassified to printer $C1150$. This is better than both the data-driven and hand-crafted feature based systems reported in literature [6].

- For classifying printers of same make and model, *CNN-$\{S^{raw}, S^{med}, S^{avg}\}_{a,e}$* gives the highest accuracies, but the misclassification in all the three methods is mainly between two printers of same make and model and it is around 7% for all three of them.

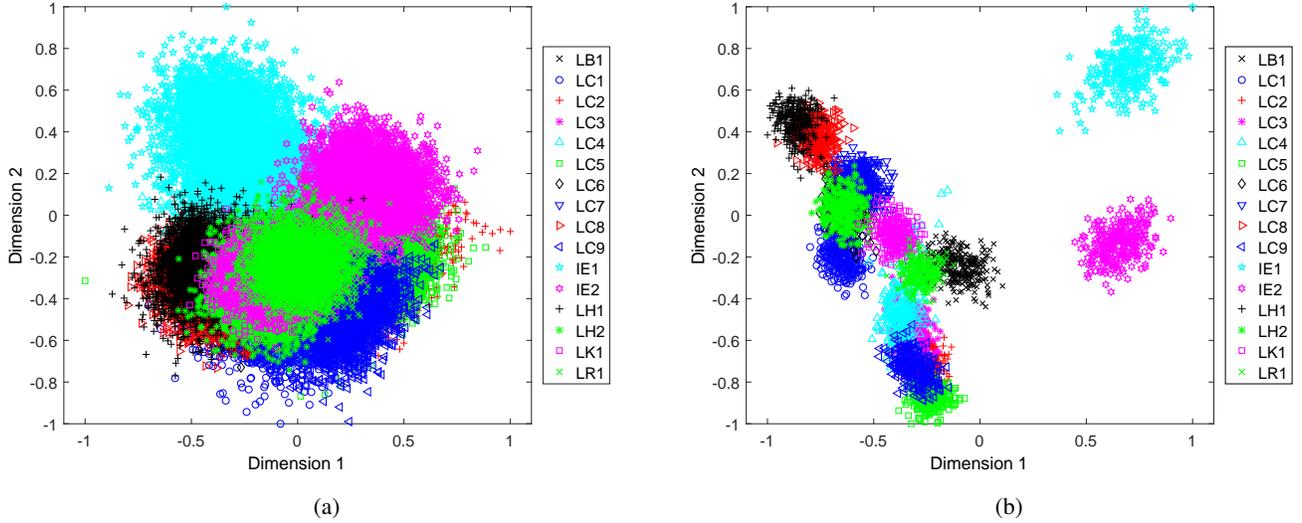

Fig. 4: Top two features of the proposed method after projecting using LDA with (a). $N_2 = 1$ and (b). $N_2 = 40$.

In addition to testing our method on the same fold as reported in [6], we also performed one hundred iterations with these random folds followed by cross-validation of these hundred iterations after an exchange of train and test data. Next, a group of first five iterations along with their corresponding cross-validated iterations are chosen resulting in a $5 \times 2$ cross-validation set. Similar sets are formed using consecutive iterations giving ninety-six such sets. Among these sets, 25 give a mean overall accuracy of 97.33% or more which is the best mean overall accuracy reported on this dataset [6]. The highest mean accuracy obtained using our $5 \times 2$ cross-validation sets is 97.68% with a standard deviation of 0.67 while the lowest average accuracy is 96.82%.

### D. Effect of Font Shapes

This set of experiments compares the performance of proposed method on pages printed with three more fonts: Arial, Times New Roman, and Comic Sans. The classifier is trained using all the letters (extracted using connected components analysis) and PoEP with $N_2 = 40$ from a single page while testing is done on all the letters (extracted using connected components analysis) from rest of the pages. Table VI shows mean and median accuracies along with standard deviations for five different choices of the training page. The results suggest that performance of the proposed method is consistently good across all four fonts.

TABLE VI: Average classification accuracy for classifying group of $N_2 = 40$ letters, using all the letters from pages scanned at 600 dpi.

| Font Type | Mean | Median | $\sigma$ |
|---|---|---|---|
| Cambria font | 99.7% | 99.7% | 0.10 |
| Arial font | 99.6 % | 99.6 % | 0.10 |
| Times New Roman font | 99.3 % | 99.5 % | 0.47 |
| Comic Sans font | 99.4 % | 99.6 % | 0.64 |

*E. Optimizing Resources using 300 dpi Scans*

All the results discussed so far correspond to documents scanned at 600 dpi. Now, the same experiments are repeated with documents scanned at 300 dpi. Similar to the previous set of experiments (Section V-D), mean and median accuracies are estimated for five iterations based on the choice of the training page. These results shown in Table VII indicate that the proposed method performs consistently well even at this resolution.

TABLE VII: Average classification accuracy for classifying group of $N_2 = 40$ letters, using all the letters from pages scanned at 300 dpi.

| Font Type | Mean | Median | $\sigma$ |
|---|---|---|---|
| Cambria font | 98.2% | 98.4% | 0.49 |
| Arial font | 99.4 % | 99.4 % | 0.13 |
| Times New Roman font | 97.8 % | 98.6 % | 2.17 |
| Comic Sans font | 98.7 % | 99.1 % | 0.72 |

*F. Printers of Same Brand and Model*

This experiment evaluates the efficacy of the proposed method for classifying printers of same brand and model. Due to the closeness of printer signatures for such printers, this intra-model scenario requires separate analysis. Here, a set of 3 printers of same brand and model are used to check the efficacy of the proposed method. As obtained from earlier experiments, the value of $N_2$ is kept fixed at 40. The average classification accuracy achieved with training data obtained from only one page per printer (around 60 feature vectors from each of the three printers) over 5 iterations is 91.9% which increases to 97.5% when training data from five pages per printer (around 280 feature vectors from each of the three printers) is used. This suggests that printers of same brand and model have significantly lower differences between their features and so larger number of training samples is required to train a classifier which can give accuracies similar to earlier experiments (Table VI). Table VIII shows the mean confusion matrix corresponding to training data obtained from five pages over ten iterations. Some groups ($N_2 = 40$) of letters are misclassified between LC10 and LC11 (refer Table I ) while LC9 achieves close to 100% accuracy. Table IX shows the average classification accuracies over five iterations (%) for classifying group of $N_2 = 40$ letters from all the 18 printers of our dataset using data from randomly selected 5 page for training, and rest of the 20 pages for testing, both in Cambria font. Corresponding average classification accuracy for all the 18 printers is 99.28%.

TABLE VIII: Confusion matrix (reporting classification accuracy in %) for classifying 3 printers of same brand and model.

| Predicted Class → True Class ↓ | LC9 | LC10 | LC11 |
|---|---|---|---|
| LC9 | 99.8 | 0.1 | 0.1 |
| LC10 | 0.1 | 96.2 | 3.7 |
| LC11 | | 3.8 | 96.2 |

## VI. CONCLUSION

A new approach for source printer classification is presented in this paper using a single classifier for all the letters of a language. As shown by experiments, this leads to a lower requirement on the amount of printed pages needed for training. For example, on our dataset of 18 printers including three printers of same make and model, only 5 pages per printer are sufficient to achieve an accuracy of more than 99%. When dealing with partially damaged test documents or documents with very few letters, our single classifier procedure would be immensely useful. At the same time, the proposed method is free of any discrepancies induced by character extraction process. By various experiments, it can be concluded that region separation along with PoEP imparts shape independence property in the proposed method which is absent in existing methods. The proposed method exhibits results comparable to CNN based existing method with experiments on existing dataset using only one sample per page to train and test the classifier. Here, our PoEP approach significantly reduces both training and testing time which could prove to be of utmost help while dealing with huge volumes of real-time data analysis in many practical scenarios.

The proposed method works consistently well for four types of fonts including Comic Sans which looks very different from the other three fonts. This again confirms the shape independence property of our method. Also, it works well even with lower resolution scanner settings which are commonly used for all practical applications as they are very fast. This could eventually pave the way for using printer source identification methods on documents scanned using smartphones (by clicking their photo). Given the high printer identification accuracy at the level of a group of letters, this method could also be used to expose various kinds of text forgeries in printed documents.

Apart from the printer artifacts, there are certainly other parameters that can impact the performance of such a system. These include aging effects on printers such as the temporal defects that may have been induced in the printer due to wear and tear over its lifetime, effect of toner ink composition and material properties of paper on which the document is printed. The performance of proposed method needs to be further tested on languages with character sets different than the English language. Given these limitations, it is clear that to make a practical framework for printed document authentication system; many techniques need to be judiciously combined.

## REFERENCES

[1] A. Roy, B. Halder, and U. Garain, "Authentication of currency notes through printing technique verification," in *Proc. Seventh Indian Conf. on Computer Vision, Graphics and Image Process.*, 2010, pp. 383–390.

TABLE IX: Results (% of correct and incorrect classification) for classifying group of $N_2 = 40$ letters (training using 5 pages and testing on 20 pages, in Cambria font).

| | LB1 | LC1 | LC2 | LC3 | LC4 | LC5 | LC6 | LC7 | LC8 | LC9 | LC10 | LC11 | IE1 | IE2 | LH1 | LH2 | LK1 | LR1 |
|---|---|---|---|---|---|---|---|---|---|---|---|---|---|---|---|---|---|---|
| Correct Classification | 100 | 100 | 99.9 | 99.9 | 99.9 | 99.7 | 99.9 | 96.6 | 99.9 | 95.3 | 96.2 | 100 | 99.8 | 100 | 100 | 100 | 100 | 100 |
| Incorrect Classification | | | (LC4, 0.1) | (LC2, 0.1) | (LC2, 0.1) | (LC2, 0.3) | (LK1, 0.1) | (LC10, 3.4) | (LC6, 0.1) | (LC10, 4.6) (LC11, 0.1) | (LC9, 3.8) | | (IE2, 0.2) | | | | | |

[2] J. Oliver and J. Chen, "Use of Signature Analysis to Discriminate Digital Printing Technologies," in *NIP & Digital Fabrication Conference*, vol. 2002, no. 1, 2002, pp. 218–222.

[3] A. K. Mikkilineni, P.-J. Chiang, G. N. Ali, G. T. C. Chiu, J. P. Allebach, and E. J. Delp III, "Printer identification based on graylevel co-occurrence features for security and forensic applications," in *Electronic Imaging*, 2005, pp. 430–440.

[4] M. J. Tsai, J. S. Yin, I. Yuadi, and J. Liu, "Digital forensics of printed source identification for Chinese characters," *Multimedia Tools and Applications*, vol. 73, no. 3, pp. 2129–2155, 2014.

[5] A. Ferreira, L. C. Navarro, G. Pinheiro, J. A. dos Santos, and A. Rocha, "Laser printer attribution: Exploring new features and beyond," *Forensic science international*, vol. 247, pp. 105–125, 2015.

[6] A. Ferreira, L. Bondi, L. Baroffio, P. Bestagini, J. Huang, J. dos Santos, S. Tubaro, and A. Rocha, "Data-driven feature characterization techniques for laser printer attribution," *IEEE Transactions on Information Forensics and Security*, 2017.

[7] O. Bulan, J. Mao, and G. Sharma, "Geometric distortion signatures for printer identification," in *Proc. IEEE Int. Conf. Acoustics, Speech and Signal Processing*, 2009, pp. 1401–1404.

[8] S. Shang, X. Kong, and X. You, "Document forgery detection using distortion mutation of geometric parameters in characters," *Journal of Electronic Imaging*, vol. 24, no. 2, pp. 023 008–023 008, 2015.

[9] J. Hao, X. Kong, and S. Shang, "Printer identification using page geometric distortion on text lines," in *IEEE China Summit and Int. Conf. Signal and Info. Process.*, 2015, pp. 856–860.

[10] Q. Zhou, Y. Yan, T. Fang, X. Luo, and Q. Chen, "Text-independent printer identification based on texture synthesis," *Multimedia Tools and Applications*, pp. 1–24, 2015.

[11] T. Ojala, M. Pietikäinen, and M. Mäenpää, "Multiresolution gray-scale and rotation invariant texture classification with local binary patterns," *IEEE Trans. pattern analysis and machine intelligence*, vol. 24, no. 7, pp. 971–987, 2002.

[12] X. Tan and B. Triggs, "Enhanced local texture feature sets for face recognition under difficult lighting conditions," *IEEE Trans. Image Process.*, vol. 19, no. 6, pp. 1635–1650, 2010.

[13] B. Zhang, Y. Gao, S. Zhao, and J. Liu, "Local Derivative Pattern versus Local Binary Pattern: Face Recognition with High-order Local Pattern Descriptor," *IEEE Trans. Image Process.*, vol. 19, no. 2, pp. 533–544, 2010.

[14] S. Murala, R. Maheshwari, and R. Balasubramanian, "Local tetra patterns: a new feature descriptor for content-based image retrieval," *IEEE Trans. Image Process.*, vol. 21, no. 5, pp. 2874–2886, 2012.

[15] G. N. Ali, A. K. Mikkilineni, E. J. Delp, J. P. Allebach, P.-J. Chiang, and G. T. Chiu, "Application of principal components analysis and gaussian mixture models to printer identification," in *NIP & Digital Fabrication Conference*, vol. 2004, no. 1. Society for Imaging Science and Technology, 2004, pp. 301–305.

[16] J.-H. Choi, D.-H. Im, H.-Y. Lee, J.-T. Oh, J.-H. Ryu, and H.-K. Lee, "Color laser printer identification by analyzing statistical features on discrete wavelet transform," in *16th IEEE International Conference on Image Processing (ICIP)*, 2009, pp. 1505–1508.

[17] W. Jiang, A. T. Ho, H. Treharne, and Y. Q. Shi, "A novel multi-size block benfords law scheme for printer identification," in *Pacific-Rim Conf. on Multimedia*. Springer, 2010, pp. 643–652.

[18] S.-J. Ryu, H.-Y. Lee, D.-H. Im, J.-H. Choi, and H.-K. Lee, "Electrophotographic printer identification by halftone texture analysis," in *IEEE International Conference on Acoustics, Speech and Signal Processing*, 2010, pp. 1846–1849.

[19] S. Elkasrawi and F. Shafait, "Printer identification using supervised learning for document forgery detection," *Proc. 11th IAPR Int. Workshop on Document Analysis Systems, DAS 2014*, pp. 146–150, 2014.

[20] T. Q. Nguyen, Y. Delignon, L. Chagas, and F. Septier, "Printer identification from micro-metric scale printing," in *IEEE Int. Conf. Acoustics, Speech and Signal Process.*, 2014, pp. 6236–6239.

[21] G. N. Ali, P.-J. Chiang, A. K. Mikkilineni, J. P. Allebach, G. T.-C. Chiu, and E. J. Delp, "Intrinsic and extrinsic signatures for information hiding and secure printing with electrophotographic devices," in *Proc. IS&T's NIP19: Int. Conf. Digital Printing Technologies*, vol. 19, 2003, pp. 511–515.

[22] A. K. Mikkilineni, G. N. Ali, P.-J. Chiang, G. T. C. Chiu, J. P. Allebach, and E. J. Delp, "Signature-embedding in printed documents for security and forensic applications," in *Proc. SPIE: Security Steganography and Watermarking of Multimedia Contents VI*, 2004, pp. 455–466.

[23] J. Picard, C. Vielhauer, and N. Thorwirth, "Towards fraud-proof id documents using multiple data hiding technologies and biometrics," in *Electronic Imaging 2004*. International Society for Optics and Photonics, 2004, pp. 416–427.

[24] P.-J. Chiang, J. P. Allebach, and G. T. C. Chiu, "Extrinsic signature embedding and detection in electrophotographic halftoned images through exposure modulation," *IEEE Trans. Inf. For. and Sec.*, vol. 6, no. 3, pp. 946–959, 2011.

[25] P.-J. Chiang, N. Khanna, A. K. Mikkilineni, M. V. O. Segovia, S. Suh, J. P. Allebach, G. T.-C. Chiu, and E. J. Delp, "Printer and scanner forensics," *IEEE Signal Process. Mag.*, vol. 26, no. 2, pp. 72–83, 2009.

[26] D.-G. Kim and H.-K. Lee, "Colour laser printer identification using halftone texture fingerprint," *Electronics Letters*, vol. 51, no. 13, pp. 981–983, 2015.

[27] H. Wu, X. Kong, and S. Shang, "A printer Forensics Method using Halftone Dot Arrangement Model," in *IEEE China Summit and International Conference on Signal and Information Processing (ChinaSIP)*, 2015, pp. 861–865.

[28] C. Schulze, M. Schreyer, A. Stahl, and T. M. Breuel, "Evaluation of graylevel-features for printing technique classification in high-throughput document management systems," in *Computational Forensics*. Springer, 2008, pp. 35–46.

[29] C. H. Lampert, L. Mei, and T. M. Breuel, "Printing technique classification for document counterfeit detection," in *Int. Conf. Computational Intelligence and Security*, vol. 1, 2007, pp. 639–644.

[30] S. Shang, N. Memon, and X. Kong, "Detecting documents forged by printing and copying," *EURASIP Journal on Advances in Signal Processing*, vol. 2014, no. 1, pp. 1–13, 2014.

[31] E. Kee and H. Farid, "Printer profiling for forensics and ballistics," in *Proc. 10th ACM workshop Multimedia and security*, 2008, pp. 3–10.

[32] O. A. Ugbeme, "Automated algorithm for the identification of artifacts in mottled and noisy images," *Journal of electronic imaging*, vol. 16, no. 3, pp. 1–11, 2007.

[33] C. Schulze, M. Schreyer, A. Stahl, and T. Breuel, "Using Dct Features for Printing Technique and Copy Detection," *Advances in Digital Forensics V*, pp. 95–106, 2009.

[34] T. Ahonen, A. Hadid, and M. Pietikäinen, "Face recognition with local binary patterns," in *European Conference on Computer Vision*. Springer, 2004, pp. 469–481.

[35] X. Wang, H. Gong, H. Zhang, B. Li, and Z. Zhuang, "Palmprint identification using boosting local binary pattern," in *IEEE Int. Conf. Pattern Recognition*, vol. 3, 2006, pp. 503–506.

[36] M. Pietikäinen, A. Hadid, G. Zhao, and T. Ahonen, *Local binary patterns for still images*. Springer, 2011.

[37] C.-C. Chang and C.-J. Lin, "LIBSVM: A Library for Support Vector Machines," *ACM Transactions on Intelligent Systems and Technology*, vol. 2, no. 3, pp. 1–27, 2011.

[38] J. Gebhardt, M. Goldstein, F. Shafait, and A. Dengel, "Document authentication using printing technique features and unsupervised anomaly detection," in *12th International Conference on Document Analysis and Recognition (ICDAR)*, 2013, pp. 479–483.

[39] Generator for randomized typographic filler text. [Online]. Available: http://generator.lorem-ipsum.info/

[40] Y.-r. Fu and S.-y. Yang, "CCS-LTP for Printer Identification based on Texture Analysis," *International Journal of Digital Content Technology & its Applications*, vol. 6, no. 13, 2012.